\documentclass[aps,preprint,preprintnumbers,superscriptaddress]{revtex4-1}

\usepackage{graphics}
\usepackage{graphicx}
\usepackage{epsf} 
\usepackage{dcolumn}
\usepackage{bm}
\usepackage{amssymb}
\usepackage{amsmath}
\usepackage{amsbsy}
\usepackage{srcltx}
\usepackage{psfrag}

\usepackage{graphicx,color,ifthen}
\usepackage{pstricks,pst-node,pst-text,pst-3d,color}

\begin{document}

\title{Ion induced solid flow}

\author{Mario Castro}
\affiliation{Grupo Interdisciplinar de Sistemas Complejos (GISC) and Grupo de Din\'amica No Lineal (DNL), Escuela T\'ecnica
Superior de Ingenier{\'\i}a (ICAI), \\ Universidad Pontificia Comillas, E-28015
Madrid, Spain}
\author{Rodolfo Cuerno}
\affiliation{Departamento de Matem\'aticas and GISC, Universidad Carlos III de Madrid, Avenida de la
Universidad 30, E-28911 Legan\'es, Spain}

\date{\today}


\begin{abstract}

Amorphous solids can flow  over very long periods of time. Solid flow can also be artificially enhanced by creating defects, as by Ion Beam Sputtering (IBS) in which collimated ions with energies in the 0.1 to 10 keV range impact a solid target, eroding its surface and inducing formation of nanometric structures. Recent experiments have challenged knowledge accumulated during the last two decades so that a basic understanding of self-organized nano-pattern formation under IBS is still lacking. We show that considering the irradiated solid to flow like a highly viscous liquids can account for the complex IBS morphological phase diagram, relegating erosion to a subsidiary role and demonstrating a controllable instance of solid flow at the nanoscale. This new perspective can allow for a full harnessing of this bottom-up route to nanostructuring.

\end{abstract}

\maketitle
\medskip

The difference between solids and liquids is straightforward for the pedestrian: fluids flow and solids don't. This distinction is not so clear cut when trying to be more quantitative. For instance, when observed over long periods of time, glaciers \cite{doake:1985} or lead pipes also flow in spite of being hard solids. Solids can indeed flow when observed at the proper time scale, the main cause being defect rearrangement. Unfortunately, experimental observation is difficult for macroscopic systems because of the long time periods involved (decades or even centuries).

\begin{figure}[!ht]
\begin{center}
\begin{psfrags}
\includegraphics[width=0.45\textwidth,clip=]{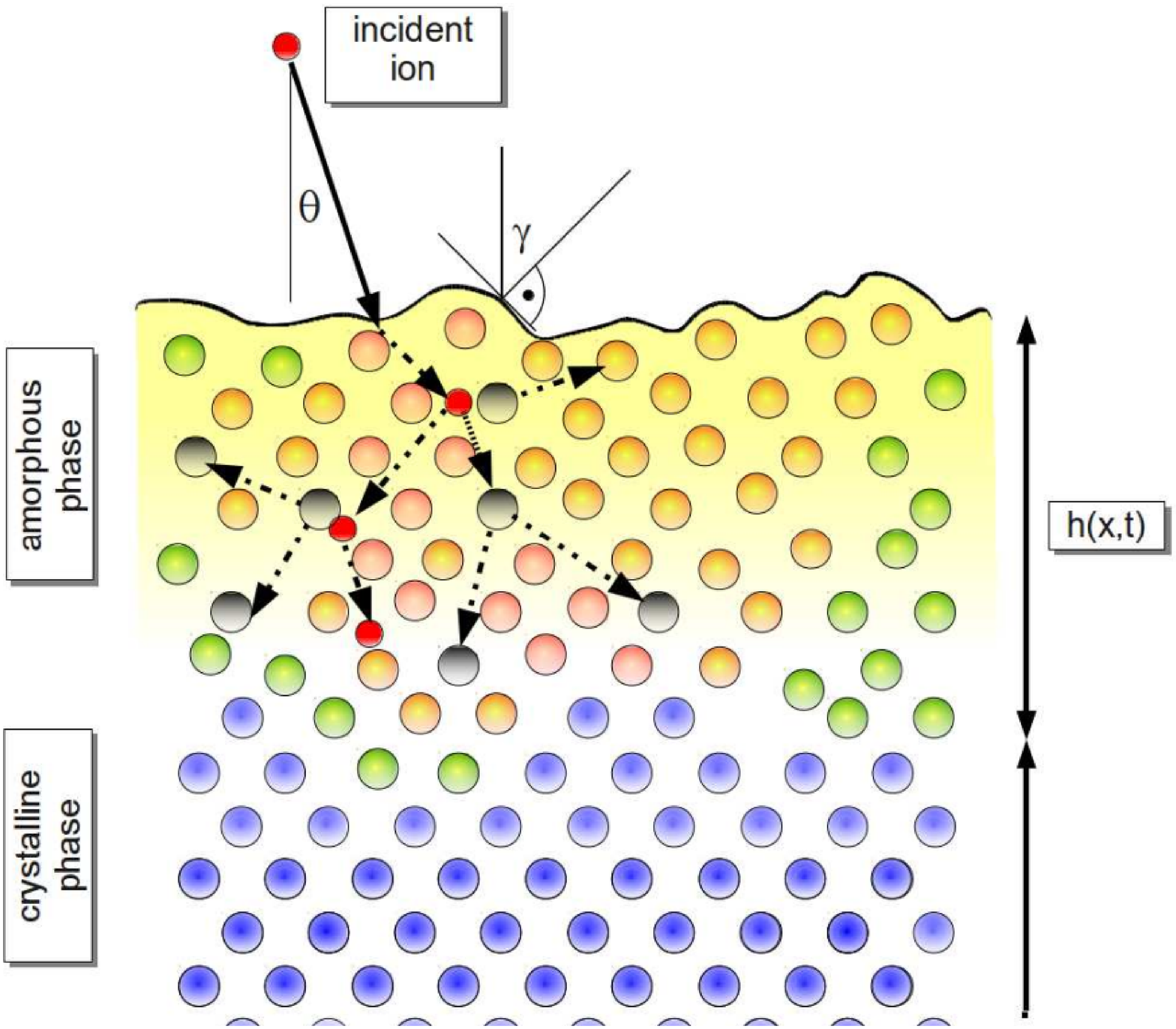}
\end{psfrags}
\end{center}
\caption{\label{cartoon} Schematic view of an IBS experiment. An incident energetic ion (red) impinges onto the target at an angle $\theta$, inducing a collision cascade that amorphizes a region of width $h(x,t)$ through the creation of vacancies and interstitials.  At long time scales the amorphous solid flows as a highly viscous fluid. Here $\gamma$ is the local slope of the surface profile.
}
\end{figure}

Solid flow at the nanoscale can be also observed or even enhanced at short times by creating artificially a large number of defects (see an sketch in Fig.\ \ref{cartoon}). Using ions with average kinetic energy $E$ from, say, 100 eV up to 10 keV, this can be achieved by Ion Beam Sputtering (IBS) of solids that become amorphous under this type of irradiation (such as semiconductors) \cite{gnaser:1998}. The ions create permanent defects in a surface layer whose thickness is of the order of the average penetration depth, trigger material ejection (sputtering) from its surface and induce material redistribution and flow \cite{volkert:1993,mayr:2003}.

Among the intriguing features of IBS, its efficiency to produce surface nanopatterns over large areas on top of a wide variety of targets, with a high degree of reproducibility \cite{facsko:1999,chan:2007_et_al}, furnishes the technique with a large potential for applications. Notably, samples that were unaffected by contamination issues have not been available until recently \cite{madi:2009,madi:2008}, a fact that has hindered a correct understanding of the mechanisms controlling this nanostructuring procedure. The ensuing experiments on Si have challenged the current paradigm according to which pattern formation is controlled by the competition between erosion processes and atom diffusion on top of the target (the so-called Bradley-Harper (BH) mechanism \cite{bradley:1988}). In fact, none of the current phenomenological models that elaborate on this view \cite{makeev:2002,facsko:2004,davidovitch:2007} are able to capture the complex morphological features elucidated by the experiments of Madi and coworkers, including the lack of pattern formation at small incidence angle $\theta$ or the existence of more than one type of morphological transitions in a $(\theta,E)$ parameter space. In these theories, the target surface is described through an effective equation in which the various physical mechanisms are combined {\em ad-hoc}, defect dynamics being neglected. The same limitations occur even for alternative models in which the evolution of the surface height is explicitly coupled to that of the density of defects whose transport is confined to a {\em thin} surface layer, akin to pattern formation on e.g.\ aeolian sand dunes, see e.g.\ \cite{notaPRLs} and references therein.

Here, we adopt a new approach to the problem by considering that the external amorphous layer produced on the target has a finite thickness and can actually be described as a highly viscous fluid. Our theory relegates the role of erosion as the dominant pattern forming mechanism and describes how solid flow can lead to both smoothening and nanostructuring, and the morphological transitions among them. This fluid-like behavior of the solid thus accounts for the pattern forming properties recently observed, our theory being validated through comparison with available experimental data. In retrospect, it is interesting to note, as in the first experimental reports half a century ago \cite{navez:1962}, that many of the appealing patterns produced by IBS resemble waves on the surfaces of water or sandy dunes (see Fig.\ 3 in \cite{appendices}). These shapes are manifestations of the physical processes occurring in the material media beneath. Thus, IBS appears as an ideal context in which a connection between fluid and solid flow can indeed be established.

In the energy range considered, an impinging ion interacts mostly with the nuclear structure of the target, generating vacancies and interstitials. Both ions and defects can diffuse at different rates, annihilate by recombination or form clusters. As borne out from Molecular Dynamics (MD) simulations, the overall effect of these processes is threefold: amorphizing \cite{pelaz:2004} and stressing the target \cite{kalyanasaundaram:2008}, and displacing the mean atom position inside the material \cite{moseler:2005}. Moreover, as also seen in experiments \cite{gnaser:1998,moore:2004,kalyanasaundaram:2008}, the amorphous layer forms rapidly, its thickness and other mechanical properties becoming stationary after a fluence of order $10^{14}$ ions cm$^{-2}$ (corresponding to a few seconds of irradiation for typical ion fluxes), the system evolves quasi-statically, and the density, $\rho$, of the target in the amorphous phase remains almost a constant. In contrast with the macroscopic solid flow of a glacier, ion fluxes employed in standard IBS experiments imply a slow driving that induces a time scale separation between atomistic relaxation rates ($\simeq 1$ ps$^{-1}$ for e.g.\ collision cascades, adatom hopping attempts), and the $\simeq 10$ s times in which significant morphological changes occur. This fact legitimates a continuum description.

In the language of fluid mechanics, we assume amorphous layer flow to be incompressible, $\nabla \cdot \mathbf{V} = 0$, where $\mathbf{V} = u \mathbf{i}+ w \mathbf{k}$ is the velocity field. Moreover, it can be considered as highly viscous because the radiation induced viscosity is around fifteen orders of magnitude larger than the viscosity of water \cite{mayr:2003}. Hence, Stokes flow applies,
\begin{equation}
\rho D\mathbf{V}/Dt = \mathbf{0} = {\bf b}+\nabla\cdot{\bf T },
\label{stokes}
\end{equation}
where $T_{ij} =-P\delta_{ij}+\mu \left(\partial_{i}u_j+\partial_j u_i\right)$ are the components of the stress tensor $\mathbf{T}$, with $P$ being pressure and $\mu$ viscosity. Eq.\ (\ref{stokes}) is supplemented with proper boundary conditions involving the erosive terms and the stress at the interfaces. Specifically, a no-slip condition is assumed at the crystalline-amorphous interface and erosion and amorphization rates are taken equal as to guarantee stationarity of the amorphous layer thickness, see \cite{appendices}.
We factor the body force term, {\bf b}, into a momentum exchange term per unit volume, times an angular contribution,
$
b=|{\bf b}|=f_E\Psi(\theta -\gamma ).
$
The coefficient $f_E$, which has units of a stress gradient, contains the coarse-grained information about the effect of the residual stress created in the target, due to ion-induced mass redistribution. 
Thus, $f_E$ 
can in principle be estimated from macroscopic measurements of stress (see below). Regarding the geometrical factor in ${\bf b}$, it only depends on the difference between the angle of incidence $\theta$ and the local surface slope $\gamma$ (see Fig.\ \ref{cartoon}). Symmetry requires $\Psi$ to be an even function of its argument. Additionally, in the limit of completely glancing incidence the ion does not collide with the target, so that $\Psi$ must vanish when $\theta -\gamma =\pi /2$. We choose $\Psi=\cos(\theta -\gamma )$ as the simplest function fulfilling these constraints. Moreover, a simple cosine has the meaning of a geometrical correction to the average ion flux.

In order to find the conditions under which the system evolves into patterns or produces smooth surfaces, we consider an initially flat surface and determine the stability of such a solution to periodic modulations of spatial wavelength $\lambda =2\pi /q$ ($q$ is the space wavenumber), whose amplitude evolves in time as $e^{\omega_q t}$. All the information about stability is contained in the dispersion relation, $\omega _q$. Hence, a modulation with wavelength $\lambda$ will grow provided the real part of the dispersion relation, $\omega^\prime _{q}$, is positive, and will be damped out for a negative $\omega^\prime _{q}$. The experimental surface will be flat only if $\omega^\prime_q <0$ for all wave vectors. The imaginary part of the dispersion relation, $\omega^{\prime\prime}_q$, contains information about how the pattern travels across the surface.
In our case, we find \cite{appendices}
\begin{equation}
\omega^\prime _q=-\frac{\left(f_E\cos(2\theta )+q^2 \sigma \right) (-2 h q+{\sinh}(2 h q))}{2q \mu  \left(1+2 h^2 q^2+{\cosh}(2 h q)\right)}+\overline{\omega}^\prime_q,
\label{rewq}
\end{equation}
where $\mu$ is the ion induced viscosity and $\sigma$ is the surface energy. For a general angular function $\Psi(\theta)$, the dispersion relation has the same form as in Eq.\ (\ref{rewq}) but replacing $\cos 2\theta $ with $\partial_\theta (\Psi(\theta )\sin\theta )$. The term $\overline{\omega}^\prime_q$ is the contribution to the dispersion relation coming from purely erosive terms. Eq.\ (\ref{rewq}) has some interesting limits in Fluid Mechanics. For instance, for $f_E\rightarrow 0$, it reduces to the classic result by Orchard \cite{orchard:1962} for the flow of a viscous layer of arbitrary thickness on top of an immobile substrate. The $q\rightarrow \infty$ limit of the latter reduces to Mullins' \cite{mullins:1959} classic rate of flattening for interface features when relaxation occurs through viscous flow $\omega _q=-\sigma q/2\mu$, already invoked in the context of IBS experiments \cite{chason:1994,frost:2009}.

\begin{figure}[!ht]
\begin{center}
\begin{psfrags}
\psfrag{angle}[][]{{\sf $\theta$ (degrees)}}
\psfrag{qnm}[][]{{\sf $q$ (nm)}}
\psfrag{wq}[][]{{\sf $\omega _q^{\prime}$ (s$^{-1}$)}}
\psfrag{energy}[][]{{\sf Energy (eV)}}
\psfrag{energy2}[][]{{\sf \footnotesize Energy (eV)}}
\psfrag{length}[][]{{\sf $\lambda $ (nm)}}
\psfrag{expthetac}[][]{{\sf \footnotesize $\theta_c^{exp}$}}
\psfrag{theothetac}[][]{{\sf \footnotesize $\theta_c^{theo}$}}
\includegraphics[width=0.59\textwidth,clip=]{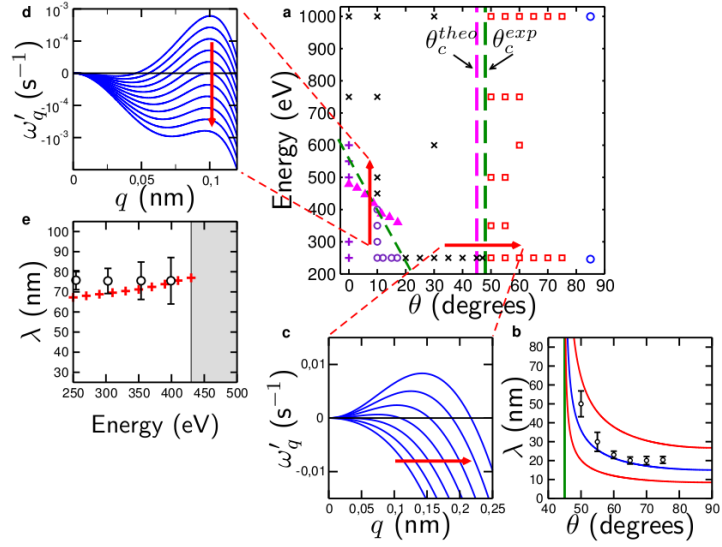}
\end{psfrags}
\end{center}
\caption{a) Experimental observations for Ar$^+$ irradiation of Si targets, taken from \cite{madi:2008,madi:2009}. Red and blue: ripple formation; black: flat surfaces; purple circles: ripple formation; purple crosses: hole formation. The dashed green lines indicate the experimental phase boundaries for a type II phase transition (vertical line at $\theta=48^{\circ}$) and for a type I transition (oblique line). The present theoretical predictions for these boundaries are shown as the purple dashed line at $\theta=\theta^{theo}_c=45^{\circ}$, and as the purple triangles, respectively. b) $\lambda_c(\theta)$ at the type II transition. Experimental data from \cite{madi:2008,madi:2009} are shown as black symbols, the green line is the experimental phase boundary, and the blue line is the present theoretical prediction, bounded within the red lines (confidence interval). c) Linear dispersion relation at type II transition. d) Linear dispersion relation at type I transition (note how the maximum of the curve gets positive at a finite value of $q$). e) Different typical wavelengths are obtained for different energies on the low energy side of this transition: Circles, experimental data from \cite{madi:2008,madi:2009}; crosses: present theoretical prediction.
}
\label{fig_nueva}
\end{figure}

If we consider solid flow as the only pattern forming mechanism, we can drop the erosive contribution $\overline{\omega}^\prime_q$, and re-evaluate this assumption {\em a-posteriori} after comparing with experiments. For the typical orders of magnitude of the layer thickness $h$ (about a few nanometers), the sign of $\omega^\prime_q$ is determined by the sign of $-\cos (2\theta)$. Thus, Eq.\ (\ref{rewq}) predicts a phase transition at $\theta_c=45^\circ$ separating smooth surfaces ($\theta <45^\circ$) from nanopattern formation ($\theta >45^\circ$). The patterns generated in the unstable region have a characteristic wavelength given by
\begin{equation}
\lambda_c =2\pi \sqrt{\frac{2\sigma}{-f_E\cos (2\theta)}}.
\label{critlambda}
\end{equation}
To be more quantitative, we need to determine the main parameters of the theory. As mentioned above, the new parameter, $f_E$, is related to the stress. There is a large experimental uncertainty in the measurement of stress, a property that depends moreover on the energy of the incident ions \cite{davis:1993,abendroth:2007}. 
Different authors report different values for ion-induced stress on Si, in the range $200$ MPa to $1.62$ GPa \cite{madi:2009,kalyanasaundaram:2008}. We are interested in the order of magnitude, so we take the geometrical mean (which accounts for the {\em average} order of magnitude) of both extreme values, thus, approximately 569 MPa. The body force is $f_{E}=$ stress/depth. Thus, the latter value of stress corresponds to $f_E=0.424$ kg nm$^{-2}$s$^{-2}$ for a layer of $3$ nm depth. In Fig.\ 2b, this value corresponds to the blue line, while the limiting values of $200$ MPa and $1.62$ GPa are used to plot the red lines, which we take as our confidence interval. More accurate measurements of stress would provide improved estimations. Additional parameters have not been measured experimentally yet, as we are proposing a new approach. However, we can provide indirect estimates for their values when these are not explicitly available, as shown in Table 1 in \cite{appendices}. With these parameters at hand, Eq.\ (\ref{critlambda}) and experimental data are seen to agree quite closely in Fig.\ \ref{fig_nueva}b.

Eq.\ (\ref{critlambda}) actually predicts power law behavior in the vicinity of the critical angle, $\lambda_c \simeq 2\pi (\sigma /f_E)^{1/2} (\theta -\theta _c)^{-1/2}$ nm, and $\omega_{q}^{\prime}\simeq f_{E}^2 h^3(3\mu \sigma)^{-1} (\theta-\theta_{c})^{2}$,
which are the same functional forms of deviation from $\theta_c$ as experimentally reported \cite{madi:2009}.
In the language of Pattern Formation \cite{madi:2008,cross:2009}, this morphological transition is a type II bifurcation, akin to an equilibrium second order phase transition, in which the typical wavelength diverges right at the transition point ($\theta=\theta_c$). Specifically, it is a continuous transition in the sense that the wave vector characterizing the pattern on the large angle side of the phase boundary emerges continuously from the origin for increasing $\theta$, as seen on Fig.\ \ref{fig_nueva}c.

An additional prediction of our theory is that the critical angle does not depend on energy, viscosity, surface tension, or on the thickness of the amorphous layer (which may contain additional angle dependences). Actually, a clear transition between $45^\circ$ and $50^\circ$ has been reported experimentally by other authors \cite{flamm:2001,ziberi:thesis} for other targets and/or irradiation conditions. The sharp theoretical transition corresponds to the purple dashed vertical line in Fig.\ 
\ref{fig_nueva}a, emphasizing the generality of our predictions in terms of parameter dependences. Thus, beyond the particular case of Si targets, our theory provides a new striking result: both the morphological transition at the critical angle and the patterns observed for more oblique incidences are exclusively caused by solid flow and not by erosion-induced instabilities. This can be better understood through an analogy: the IBS system is similar to a liquid flowing down an inclined plane \cite{craster:2009} with inclination angle $\theta$ (see Fig.\ 2 in \cite{appendices}). The only difference between both systems is that, here, the external body force, ${\bf b}$, is not constant (as gravity is in the fluid analog) but depends, rather, on the local inclination angle (through the dependence of $\Psi$ on $\gamma $). Thus, in the IBS case the effective gravity is larger in those locations where the surface {\em faces} the external beam. Precisely, for $\theta>\theta _c$, the net difference between the force on both sides of an undulation makes it shrink laterally and its amplitude grows due to incompressibility, this being the physical origin of the instability.

Going back to the experiments summarized in Fig.\ \ref{fig_nueva}, there are additional important morphological features in the $(\theta,E)$ diagram besides the transition we have just discussed. Namely, there is another transition at small angles and low energies (indicated by the oblique green dashed line in the figure). This transition is of type I, namely, the typical wavelength remains finite at the transition, akin to an equilibrium first order phase transition \cite{madi:2008,cross:2009} in which the wave vector characterizing the system jumps discontinuously from $q=0$ to a finite value. Given that the solid flow mechanism is morphologically stable below $\theta_c$, a destabilizing mechanism must prevail at low incidence angles in order to observe a patterned structure. A natural candidate is erosion, so far assumed to be the only cause of morphological instabilities in these systems \cite{chan:2007_et_al}. We thus need to quantify the balance between the erosion and flow mechanisms by choosing a specific form of the erosive part of the dispersion relation, $\overline{\omega}^{\prime}_q$. Its classic form is \cite{chan:2007_et_al}
\begin{equation}
\overline{\omega}^\prime_q=-\nu^{BH}_E(\theta) q^2-{\mathcal K}q^4.
\label{BHlike}
\end{equation}
where $\nu^{BH}_E(\theta)$ is {\em negative} at small incidence angles. Given this choice, we show in Fig.\ \ref{fig_nueva}d
the shape of the full dispersion relation $\omega^{\prime}_q$ for increasing energies at a fixed small incidence angle, in such a way that the experimental phase boundary is crossed. 
Indeed, the less stable wave vector jumps discontinuously to zero from a finite value at the positive maximum of $\omega^{\prime}_q$, as expected for a type I transition. Also, the length scale of the pattern on the low energy side of the transition is in good agreement with the experimental one (Fig.\ \ref{fig_nueva}e), 
and, as the latter, shows a weak dependence on the ion energy.

Note that the BH estimate that we are using for $\nu^{BH}_E(\theta)$ is known from simulations, using e.g.\ the binary collision approximation, not to be very accurate numerically \cite{chan:2007_et_al}.
This may be the reason why the shape of the transition curve (purple triangles in Fig.\ \ref{fig_nueva}a) does not agree with the experimental one. However, the main conclusion is that BH type mechanisms can still compete with solid flow and induce instabilities at small incidence angles, provided the average ion energy is sufficiently small.

Returning to the main questions considered in the introduction, ion induced solid flow allows to understand the intricate variety of morphologies observed in the laboratory at low to intermediate ion energies, and the transitions among them. In this way hydrodynamic interactions may constitute the previously hypothesized non-local effects \cite{madi:2008,davidovitch:2007} required to account in particular for the observed type I transition. Interestingly, to some extent this fact parallels the case at higher ion energies ($\simeq$ 1 MeV), although strong differences persist, like the predominance for the latter of electronic scattering over the nuclear contribution to the stopping power, leading to strong many body effects over binary collision cascades, and the ensuing viscoelastic flow \cite{trinkaus:1995}.

At any rate, one can exploit the analogy with fluid flow to translate both the theoretical toolbox and the physical insight to understand more profoundly the complex interplay between relaxation mechanisms in IBS. Intuitively, the vision of the solid as a slowly moving fluid on whose surface nanowaves can be created, can drastically impact the controlled production of nanostructures in semiconductors. Yet, our approach also raises many fundamental questions and predictions whose assessment requires new atomistic simulations and experiments.
Thus, many details need to be quantitatively determined, in a similar fashion as viscosity had to be tabulated in the past, following application of Newton's laws to Fluid Dynamics. This line of research opens new theoretical and practical issues to be elucidated in the coming years.

\appendix

\section{Constitutive equations and boundary conditions}

The equations for the conservation of mass,
\begin{equation}
\nabla\cdot{\bf V}=0,
\label{continuity}
\end{equation}
and momentum
\begin{equation}
0={\bf b}+\nabla\cdot{\bf T },
\label{momentum}
\end{equation}
introduced in the main text, need to be supplemented with proper constitutive equations and boundary conditions. We will assume the solid to flow as a Newtonian fluid, thus the stress tensor is given by
\begin{equation}
T_{ij} =-P\delta_{ij}+\mu \left(\partial_{i}u_j+\partial_j u_i\right),
\label{tau}
\end{equation}
with $i,j=\{x,y\}$, where $P$ is hydrostatic pressure (that is an output of the solution of the full problem) and $\mu$ is viscosity. We consider ions that are impigning in planes that are parallel to the $(x,y)$ plane, and assume translational invariance in the perpendicular direction for simplicity. Given the fast stabilization of the thickness of the amorphous layer seen in experiments \cite{gnaser:1998} and in MD simulations \cite{kalyanasaundaram:2008,moore:2004},
we assume that the amorphous layer is in a stationary state in which it is bounded between two interfaces. Denoting by $h^{(a)}$ the position of the free target surface above a reference plane and by $h^{(c)}$ the position of the crystalline-amorphous interface that exists beneath (so that $h(x,t) = h^{(a)}-h^{(c)}$ in Fig.\ 1 of the main text), we assume a no-slip boundary condition at the crystalline-amorphous interface, while we request a balance of stress at the free interface along the local normal direction
\begin{equation}
\frac{T_{xx}(h_x^{(a)})^2+T_{zz}-2h^{(a)}_xT_{xz}}{1+(h_x^{(a)})^2}=\frac{\sigma h^{(a)}_{xx}}{(1+(h_x^{(a)})^2)^{3/2}}+\Pi ,
\label{surface}
\end{equation}
and in the tangential direction
\begin{equation}
\frac{(T_{zz}-T_{xx})h_x^{(a)}+(1-(h^{(a)}_x)^2)T_{xz}}{1+(h_x^{(a)})^2}=\frac{d\sigma }{ds}+\tau .
\label{surface2}
\end{equation}
Here, $\Pi$ and $\tau $ are external normal and shear stresses applied directly on the surface, that will be henceforth neglected, $\sigma $ is the surface tension of the amorphous-vacuum interface, and $s$ is arc-length. Explicit non-zero values for these stresses could be used for further improvements of the theory ---for instance, one could consider contributions to $\Pi$ and $\tau$ that are induced by implanted ions in the near-surface region.

Finally, the kinematic condition that the material derivative is zero at each interface leads to evolution equations for both $h^{(a)}$ and $h^{(c)}$ as
\begin{equation}
w-\partial_t h^{(c)}-u\partial_x h^{(c)}=j_{am},
\label{crystal}
\end{equation}
\begin{equation}
w-\partial_t h^{(a)}-u\partial_x h^{(a)}=j_{er} ,
\label{erosionsurface}
\end{equation}
where $j_{er}$ and $j_{am}$ are rates of erosion and amorphization, respectively. For instance, the erosive (BH-type) dispersion relation enters directly through $j_{er}$, so that its contribution to the full dispersion relation is additive, as used in the main text. Again, due to the assumed stationarity of the amorphous layer thickness, we will take both rates to be equal on average. Moreover, for simplicity, we assume that the crystalline-amorphous interface is flat, since it can be shown that deviations from flatness only influence the imaginary part of the dispersion relation, while only the real part influences ripple formation and its stability.

Finally, the linear dispersion relation is obtained by solving the linearized version of the previous set of equations for a periodic perturbation of the flat solution, $h(x,t) = e^{\omega _qt-iqx}$, where $h = h^{(a)}-h^{(c)}$ following the standard procedure (for details, see e.g.\ Ref. \cite{cross:2009}).

\section{Predictions of the theory in the proximities of the type II transition}

\begin{figure}[!htp]
\begin{center}
\begin{psfrags}
\psfrag{theta0}[][]{$\theta<45^\textrm{o}$}
\psfrag{thetacasi45}[][]{$\theta\geq 45^\textrm{o}$}
\psfrag{thetamayor45}[][]{$\theta>45^\textrm{o}$}
\psfrag{thetamuchomayor}[][]{$\theta\gg 45^\textrm{o}$}
\includegraphics[width=0.49\textwidth,clip=]{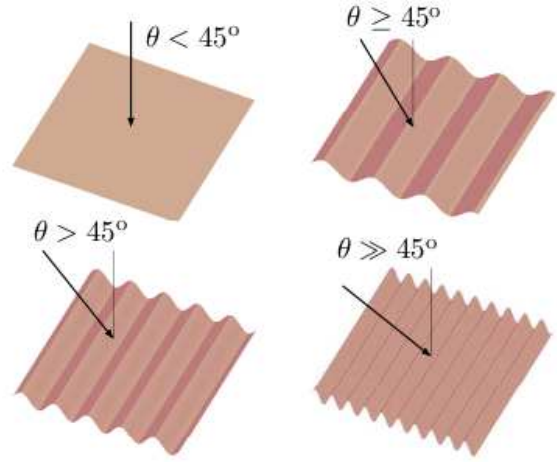}
\end{psfrags}
\end{center}
\caption{The surface morphology changes with the incidence angle. This cartoon illustrates the transition at $\theta^{theor}_c=45^\textrm{o}$ from {\em flat} to {\em patterned} surfaces. The solid black arrow indicates the direction of the ion beam. Above the critical angle, the characteristic size of the pattern is smaller for larger angles, according to $\lambda_c =2\pi [2\sigma/(-f_E\cos 2\theta)]^{1/2}$.}
\label{cartoon2}
\end{figure}

The typical length scale of the pattern above $\theta_c=45^{\circ}$ is given by
\begin{equation}
\lambda=2\pi\sqrt{\frac{2\sigma}{-f_{E}\cos(2\theta)}}.
\end{equation}
Expanding the cosine in the proximity of the critical angle, we find
\begin{equation}
\cos(2\theta)=\cos(2\theta_{c})-2\sin(2\theta_{c})(\theta-\theta_{c})+\cdots\simeq-2(\theta-\theta_{c}),
\end{equation}
so that
\begin{equation}
\lambda\simeq 2\pi\sqrt{\frac{\sigma}{f_{E}}}(\theta-\theta_{c})^{-1/2},
\end{equation}
as observed in the experiments of Ref.\ \cite{madi:2009}. Using this approximation in the full dispersion relation and assuming that $h/\lambda\ll 1$ we find that
\begin{equation}
\omega_{q}^{\prime}\simeq \frac{f_{E}^2 h^3}{3\mu \sigma} (\theta-\theta_{c})^{2},
\end{equation}
also in agreement with the experiments. Eqs.\ (10)-(11) are as quoted in the main text. As stressed there, the critical angle depends only on the form of $\Psi(\theta-\gamma)$ so that the numerical values of $h$, $\mu$, $\sigma$, etc.\ do not affect the predicted value of $\theta_{c}$. In Fig.\ \ref{cartoon2} we show, schematically, how the ensuing surface morphology can be tuned by changing the incidence angle.

\section{Numerical values of the parameters}

\begin{table}[!h]
\begin{center}
\begin{tabular}{|c|c|c|}\hline
{\bf Parameter} &{\bf Value}&{\bf Reference}\\ \hline
Surface tension ($\sigma$)&$1.34$ J/m$^{2}$& \cite{vauth:2007} \\ \hline
Amorphous layer thickness ($250$ eV) &$3$ nm& \cite{madi:2008,madi:2009} \\\hline
Viscosity ($\mu$)& $6\times 10^8$ Pa s & \cite{madiMRS}$^{*}$\\ \hline
Stress range&$200$ MPa--$1.62$ GPa&\cite{madi:2009,kalyanasaundaram:2008}\\ \hline
\end{tabular}
\caption{Numerical values of the parameters used in the main text. $^{*}$Mayr et al.\ \cite{mayr:2003} report the value $\mu=10^{9}$ Pa s, which also leads to good agreement with the results presented in this work.}
\end{center}
\end{table}

The type I transition at low angles and small energies arises from the competition between the stabilizing solid flow (through parameter $f_{E}$) and a destabilizing mechanism. Until more refined measurements can be done at these incidence angles, we postulate erosion (sputtering) as described by BH as a natural candidate. We obtain estimates of the erosion rates from the TRIM software \cite{trim}. Thus, for Si at normal incidence \cite{adrian}, $\nu_{E}=3.6$ nm$^{2}$s$^{-1}$. The dependence of this value with energy is linear \cite{makeev:2002}. Similarly, the amorphous layer width scales with energy as $h\sim E^{2m}$ with $m\simeq 0.31$ \cite{ziberi:thesis} for the range of energies considered. On the other hand, $f_{E}$ is related to stress and to the amorphous layer thickness. In Ref.\ \cite{davis:1993} the connection between stress, $\Sigma$, and energy is given by
\begin{equation}
\Sigma \propto \frac{E^{1/2}}{r + k E^{5/3}},
\end{equation}
where $r=1.7$ and $k=0.0002$ are taken from \cite{taylor:2009} for C (these parameters seem not to be available for Si). With these values, we use the constant ${\cal K}$ in Eq.\ (4) of the main text as a free parameter, finding ${\cal K}=38.2$ nm$^{4}$s$^{-1}$ in order to have a characteristic wavelength around $75$ nm. Estimates for other parameters employed in the main text appear on Table 1.

The lack of precise parameter values for Si introduces uncertainties in the quantitative value of the dispersion relation. However, the qualitative form of the full linear dispersion relation $\omega^{\prime}_q$ seems to be valid, and stresses the non-local character of our theory as a main ingredient to account for the Type I transition. We expect further developments in simulation techniques as well as more accurate experiments to improve the present predictions.

\section{Analogy with a fluid flowing down an inclined plane}

\begin{figure}[!htp]
\begin{center}
\begin{psfrags}
\includegraphics[width=0.9\textwidth,clip=]{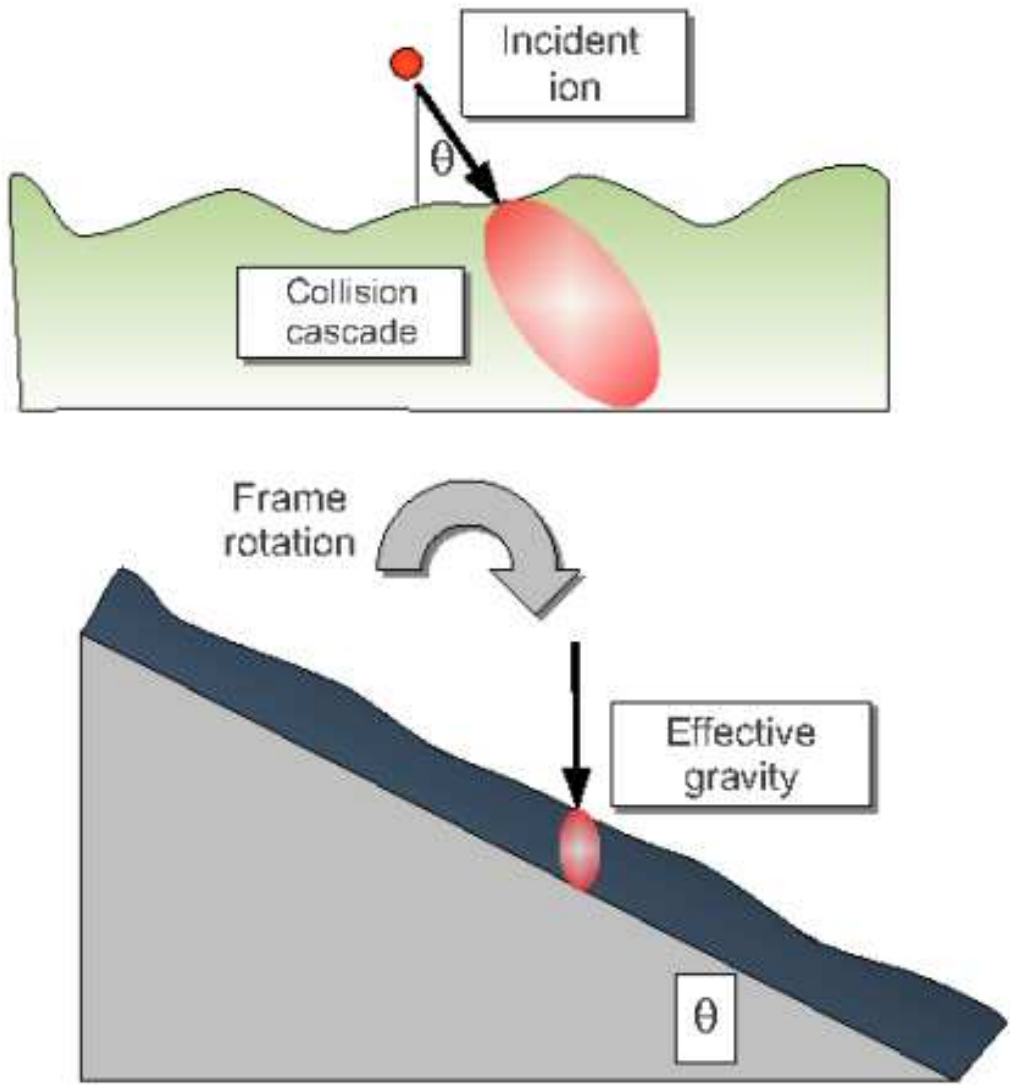}
\end{psfrags}
\end{center}
\caption{\label{analogy}Analogy between IBS and fluid flow down a inclined plane. By rotating the frame of reference so that the ion impacts vertically, the IBS system can be seen as fluid flow down an inclined plane. The body force {\bf b} is equivalent to a non-homogeneous gravity field that depends on the local slope value at the free interface.}
\end{figure}

In Fig.\ \ref{analogy} we provide a cartoon to illustrate the analogy between ion induced solid flow and a genuine fluid flowing down an inclined plane. The two systems are mathematically equivalent once a rotation of angle $\theta$ is performed in the laboratory reference frame. The main difference between both systems comes from the corresponding gravity fields. While for the standard fluid problem gravity is a constant, for the erosion system the effective gravity changes from point to point on the surface, its value depending strongly on the local surface orientation with respect to the ion beam. Recalling the assumption on fluid incompressibility that has been made, this fact accounts physically for the morphological instability that appears at $45^\circ$, as discussed in the main text.
%

\begin{acknowledgments}

We thank M.\ Aziz, S.\ Norris, C.\ Madi, and M.\ P.\ Brenner for discussions and comments.
This work has been partially supported by MICINN (Spain) Grants No.\ FIS2009-12964-C05-01 and No.\ FIS2009-12964-C05-03. 
\end{acknowledgments}


\begin{figure}[!ht]
\begin{center}
\begin{psfrags}
\includegraphics[width=0.45\textwidth,clip=]{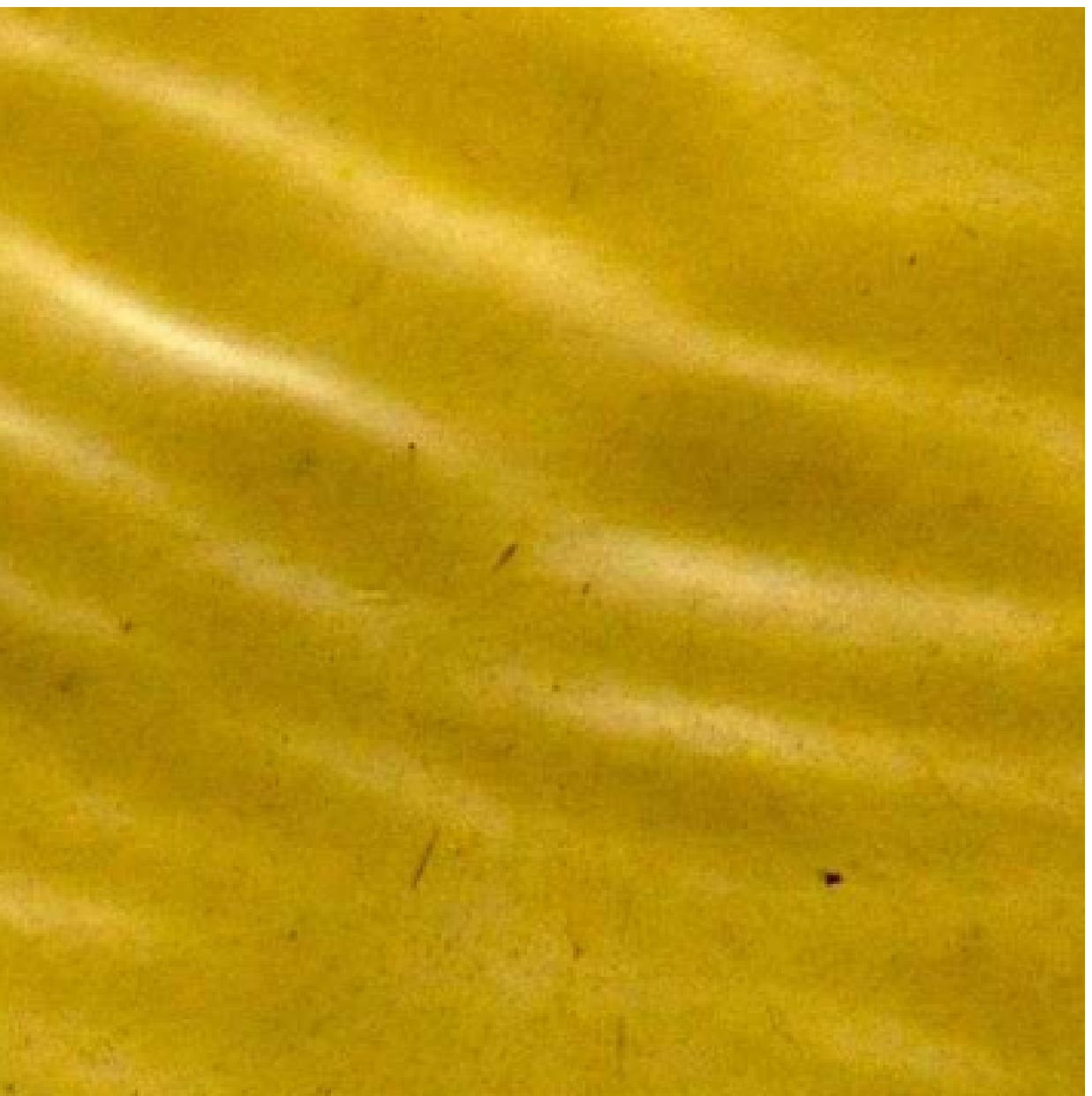}\\
\includegraphics[width=0.45\textwidth,clip=]{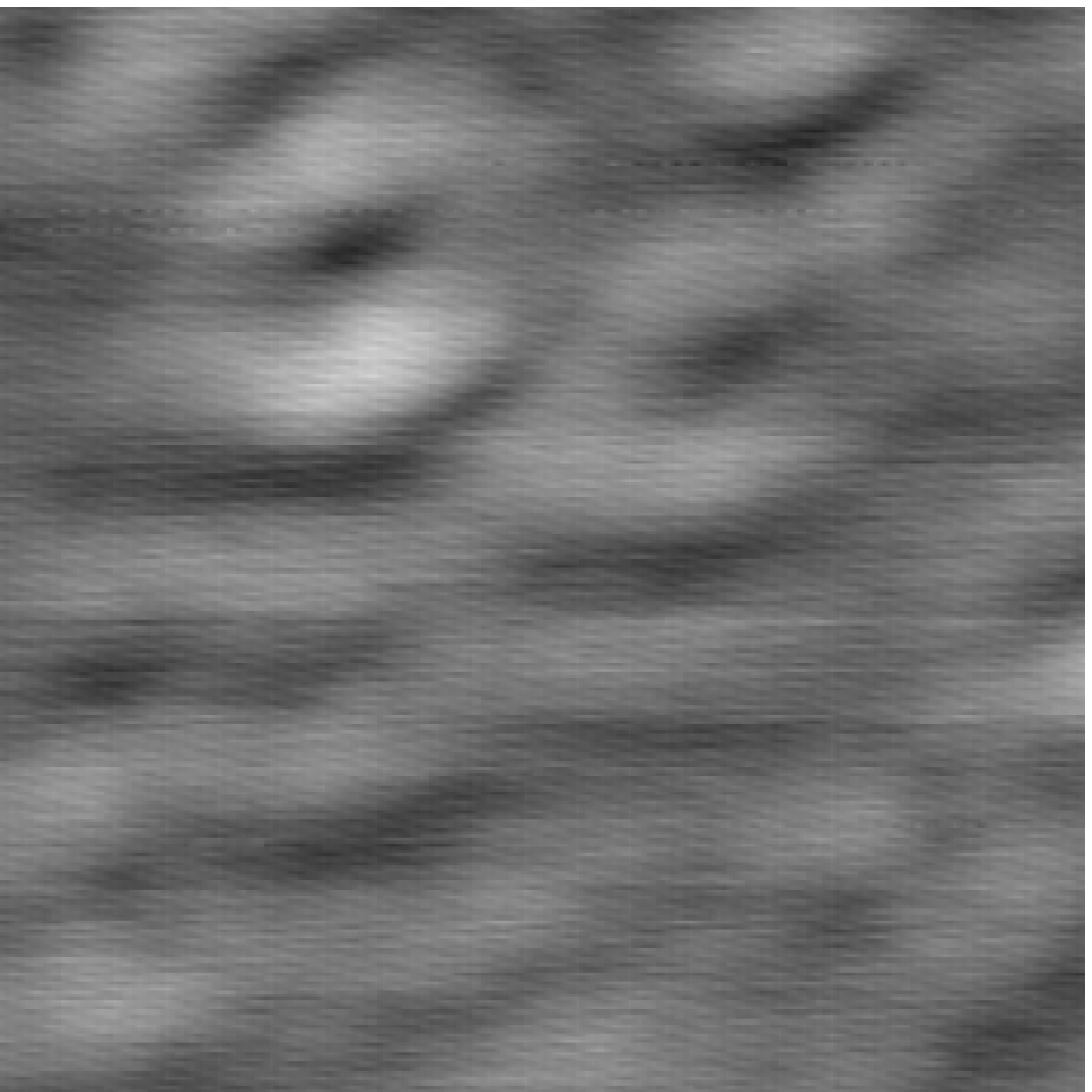}
\end{psfrags}
\end{center}
\caption{\label{ripples} Water ripples (top) of characteristic wavelength $1$ cm vs Ion Beam Sputtering (IBS) ripples (bottom) on Si. Water ripples were obtained by shaking a photography bucket filled with water (courtesy of R. Vida, Universidad Pontificia Comillas). Silicon ripples were obtained by irradiation with Ar$^+$. The width of the observation window is $1$ $\mu$m$^2$ (courtesy of L. V\'azquez, Instituto de Ciencia de Materiales de Madrid-CSIC). The naked eye similarity between these macroscopic and microscopic patterns were already noted in early works on IBS \cite{navez:1962}.}
\end{figure}


\begin{thebibliography}{99}
\bibitem{appendices} See appendices with details on our derivations, parameter estimation, and auxiliary figures.
\end{thebibliography}
\end{document}